\documentclass{pasj00}
\begin{document}
\SetRunningHead{S.\ Yamauchi et al.}{
Iron Emission Lines on the Galactic Ridge Observed with Suzaku}
\Received{}%{yyyy/mm/dd}
\Accepted{}%{yyyy/mm/dd}

\title{Iron Emission Lines on the Galactic Ridge Observed with Suzaku}

%%% begin:list of authors
 \author{%
   Shigeo \textsc{Yamauchi},\altaffilmark{1}
   Ken \textsc{Ebisawa},\altaffilmark{2}
   Yasuo \textsc{Tanaka},\altaffilmark{3}  
   Katsuji \textsc{Koyama},\altaffilmark{4} 
   Hironori \textsc{Matsumoto},\altaffilmark{4} \\
   Noriko Y. \textsc{Yamasaki},\altaffilmark{2}
   Hiromitsu \textsc{Takahashi},\altaffilmark{5} and 
   Yuichiro \textsc{Ezoe},\altaffilmark{6}
}
 \altaffiltext{1}{Faculty of Humanities and Social Sciences, Iwate University,  
3-18-34 Ueda, Morioka, Iwate 020-8550}
 \email{yamauchi@iwate-u.ac.jp}
 \altaffiltext{2}{Institute of Space and Astronautical Science/JAXA, 
  3-1-1 Yoshinodai, Sagamihara, Kanagawa 229-8510}
 \altaffiltext{3}{Max-Plank-Institut f\"ur extraterrestrische Physik,
  D-85748 Garching, Germany}
 \altaffiltext{4}{Department of Physics, Graduate School of Science, 
  Kyoto University, Sakyo-ku, Kyoto, 606-8502}
 \altaffiltext{5}{Department of Physical Science, School of Science, 
  Hiroshima University, \\ 1-3-1 Kagamiyama, Higashi-Hiroshima, Hiroshima, 
739-8526}
 \altaffiltext{6}{Department of Physics, 
  Tokyo Metropolitan University, 1-1 Minami-Osawa, Hachioji, Tokyo 192-0397}

%%% end:list of authors

%% `\KeyWords{}' always has to be placed before `\maketitle'.
 %Do NOT move this preamble from here!

\maketitle

\begin{abstract}
In order to elucidate origin of the 
Galactic Ridge X-ray Emission,
we analyzed Suzaku data taken at various regions along the 
Galactic plane and 
studied their Fe-K emission line features. 
Suzaku resolved the Fe line complex into three narrow lines at $\sim$6.4 keV, 
$\sim$6.7 keV and $\sim$6.97 keV, which are K-lines from neutral 
(or low-ionized), 
He-like, and H-like iron ions, respectively. 
The 6.7 keV line is clearly seen in all the observed regions and its
longitudinal distribution is consistent with 
that determined from previous observations.
The 6.4 keV emission line was also found in various
Galactic plane regions ($b\sim0^{\circ}$).
Differences in flux ratios of the 6.4 keV/6.7 keV 
and 6.97 keV/6.7 keV lines 
between the Galactic plane and the Galactic center regions are 
studied and its implication is discussed.
\end{abstract}

\section{Introduction}

The Galactic Ridge X-ray Emission (GRXE) is unresolved X-ray emission along 
the Galactic plane
(e.g., 
\cite{Worrall1982,Warwick1985,Koyama1986,Yamauchi1996,Kaneda1997,Sugizaki2001,
Ebisawa2001,Tanaka2002,Revnivtsev2006}).
Previous observations with several X-ray satellites have revealed 
spectral characteristics of the GRXE.
The GRXE spectrum exhibits emission lines from highly ionized Si, S and Fe, and
 the 3--10 keV continuum spectrum is 
represented by a thin thermal emission model
with a temperature of $\sim$5--10 keV \citep{Koyama1986,Kaneda1997,
Sugizaki2001}.
Additionally, a significant excess from the thermal emission is found 
above 10 keV \citep{Yamasaki1997,Valinia1998}.

Spatial distribution  of the GRXE is very similar to the Galactic structure 
\citep{Warwick1985,Yamauchi1993}, and traces 
near infrared emission \citep{Revnivtsev2006}.
Fluctuation analysis using the ASCA data has given a constraint on
the X-ray point sources contributing to the GRXE:
the number density is larger than 110 (3$'\times$3$'$)$^{-1}$ 
and the averaged flux is smaller than 10$^{-14}$ erg s$^{-1}$ cm$^{-2}$
\citep{Sugizaki1999}.
If the unresolved faint X-ray sources are located in the 4 kpc arm, 
number of the sources in the Galaxy ($N$) and their luminosity 
($L_{\rm X}$) should be  $N>$1.4$\times$10$^{7}$ 
and $L_{\rm X}<1.9\times10^{31}$ erg s$^{-1}$ (2--10 keV), respectively 
\citep{Sugizaki1999}.
Meanwhile, a deep observation with Chandra %with a superior spatial resolution 
resolved only $\sim$10 \% or $\sim19 \%$ of the GRXE flux  
into discrete sources with 
an X-ray flux ($F_{\rm X}$) of
$F_{\rm X}>3\times10^{-15}$ erg s$^{-1}$ cm$^{-2}$ 
in the 2--10 keV band (Ebisawa et al. 2001, 2005)
or 
$F_{\rm X}>1.2\times10^{-15}$ erg s$^{-1}$ cm$^{-2}$ 
in the 1--7 keV band \citep{Revnivtsev2007}, respectively.

While our observational knowledge on the GRXE has significantly
increased since its discovery, origin of the GRXE remains unsolved.
The most important issue is whether the GRXE is truly diffuse emission or 
composition of numerous faint X-ray sources.
However, in either way,  problems still remain:
If the diffuse origin is right, the hot plasma gas 
having a huge thermal energy ($\sim10^{56}$ erg, a filling factor = 1) 
is hard to be confined by the Galactic gravity. 
What is the mechanism to produce and maintain 
such a hot plasma gas in the Galaxy? 
On the other hand, if the point source model is correct, 
what kind of sources are they? % which we have yet to see?
The sources are required to have a thin thermal emission 
($kT$=5--10 keV) with  strong Fe-K emission lines, 
similar to the GRXE.

\begin{table*}[htb]
\begin{center}
\caption{Observation logs.}
\begin{tabular}{lcccc} \hline \\ [-6pt]
Position ID & Sequence No. & ($l$, $b$) & Observation time (UT)& Exposure \\ 
         &              &            &  Start~~~~--~~~~End                & (ksec)    \\ 
\hline\\ [-6pt]
\multicolumn{5}{c}{Galactic Ridge X-Ray Emission (GRXE)}\\
\hline\\ [-6pt]
R1 & 100028020 & (\timeform{332.D00}, \timeform{-0.D15}) & 2005-09-18 22:36 -- 2005-09-19 11:58 & 18\\  
R2 & 100028010 & (\timeform{332.D40}, \timeform{-0.D15}) & 2005-09-19 11:58 -- 2005-09-20 19:38 & 40\\  
R3 & 100028030 & (\timeform{332.D70}, \timeform{-0.D15}) & 2005-09-20 19:38 -- 2005-09-21 07:29 & 21\\  
R4 & 100026030 & (\timeform{345.D80}, \timeform{-0.D54}) & 2005-09-28 07:06 -- 2005-09-29 04:25 & 37\\  
R5 & 100026020 & (\timeform{347.D63}, \timeform{0.D71})  & 2005-09-25 19:02 -- 2005-09-26 15:42 & 29\\  
R6 & 500008010 & (\timeform{8.D04}, \timeform{-0.D05})   & 2006-04-07 11:48 -- 2006-04-08 10:54 & 40\\  
R7 & 500007010 & (\timeform{8.D44}, \timeform{-0.D05})   & 2006-04-06 14:13 -- 2006-04-07 11:48 & 37\\  
R8 & 500009010 & (\timeform{28.D46}, \timeform{-0.D20})  & 2005-10-28 02:24 -- 2005-10-30 21:30 & 83\\  
\hline\\ [-6pt]
\multicolumn{5}{c}{Galactic Center Diffuse X-Ray Emission (GCDX)}\\
\hline\\ [-6pt]
GC1 & 500019010 & (\timeform{358.D91}, \timeform{-0.D04}) & 2006-02-23 10:50 -- 2006-02-23 20:02 & 13\\  
GC2a& 100027020 & (\timeform{359.D75}, \timeform{-0.D05}) & 2005-09-24 14:16 -- 2005-09-25 17:27 & 37\\  
GC2b& 100037010 & (\timeform{359.D75}, \timeform{-0.D05}) & 2005-09-29 04:25 -- 2005-09-30 04:29 & 43\\  
GC3a& 100027010 & (\timeform{0.D06}, \timeform{-0.D07})   & 2005-09-23 07:07 -- 2005-09-24 11:05 & 44\\  
GC3b& 100037040 & (\timeform{0.D06}, \timeform{-0.D07})   & 2005-09-30 07:41 -- 2005-10-01 06:21 & 42\\  
GC4 & 100037070 & (\timeform{1.D00}, \timeform{-0.D10})   & 2005-10-12 07:05 -- 2005-10-12 11:05 & 9\\  
%%%\hline\\ [-6pt]
%%%B1  & 500020010 & (\timeform{1.D30}, \timeform{-3.D50})   & 2006-03-07 17:56 -- 2006-03-08 01:39 & 13\\  
\hline \\ 
\end{tabular}
\end{center}
\end{table*}
%\vspace{-0.5cm}

Systematic study of the GRXE spectra 
is important to solve origin of the GRXE. 
Suzaku has better spectral
resolution, wider energy band, and lower/more stable intrinsic background than 
previous X-ray satellites (Mitsuda et al. 2007). 
Therefore, Suzaku is the best instrument to study 
diffuse sources with low surface brightness such as the GRXE.
In fact, the 100 ks observation in the Scutum region of 
($l$, $b$)=(\timeform{28.D46}, \timeform{-0.D20}) 
with Suzaku for the first time
revealed that the Fe line complex consists of three narrow emission lines,
which are K-lines from neutral or low ionized (6.4 keV), He-like (6.68 keV), 
and H-like (6.97 keV) irons \citep{Ebisawa2008}.
In particular, existence of the 6.4 keV line on the Galactic plane 
raises a new question on its origin, 
because the 6.4 keV line is not expected from a 
thin thermal plasma.
It has not yet been known 
whether the spectral properties of the GRXE, 
in particular the Fe-line features, in all of the Galactic plane regions 
are the same or not.
In order to examine spectral and spatial variations of the GRXE,
we are going to analyze the GRXE spectra 
obtained with Suzaku in various regions on the Galactic plane and in the 
Galactic center (GC) region 
and compare their iron line features.

\section{Observations and Data Reduction}

Suzaku observations of the Galactic plane were
carried out with 4 X-ray CCD camera systems (XIS, \cite{Koyama2007a}) 
placed at the focal plane 
of the thin-foil X-ray Telescopes (XRT, \cite{Serlemitsos2007}),
and with the co-aligned non-imaging 
Hard X-ray Detector (HXD, \cite{Takahashi2007, Kokubun2007}).
In order to compare the spectral features of the thermal component, 
especially iron line features of the GRXE, 
we concentrate on the XIS data analysis in this paper.

XIS sensor-1 (XIS1) has a back-illuminated CCD (BI), while
the other three XIS sensors (XIS0, 2, and 3) 
have  front-illuminated CCDs (FI).
The XIS was operated in the normal clocking mode. 
The field of view (FOV) of the XIS is 17.8$'$$\times$17.8$'$.
The non-X-ray background level of the FI detectors 
is lower than that of the BI detector, and
the sensitivity of the FI detector in the Fe band 
is better than that of the BI detectors \citep{Koyama2007a}. 
Therefore, in order to achieve the highest signal-to-noise ratio 
in the Fe band, we used only the FI detectors.

We selected the Galactic plane fields at $| b |<$1$^{\circ}$ 
which are devoid of
significant point sources (position ID: R1--8).
Galactic locations and observation dates of the data used 
for the present analysis
are listed in table 1.
We note that the data of R8 are the same as those used in \citet{Ebisawa2008}.
We also analyzed the data of R8 in the same process.
In order to compare the iron line features of the 
GRXE with those of the Galactic 
center diffuse X-ray emission (GCDX) \citep{Koyama1989,Yamauchi1990},
we also selected  data observed near the GC
(position ID: GC1--4, see table 1).

Data reduction and analysis were made using the HEADAS software 
version 6.2 and version 1.2 of the processed data.
We excluded the data obtained at the South
Atlantic Anomaly, during Earth occultation, and at  low elevation
angles from the Earth rim of $<$ 5$^{\circ}$ (night Earth) or $<20^{\circ}$
(day Earth). 
We also removed the XIS hot and/or  flickering pixels.
The resultant exposure time is also listed in table 1.

\section{Analysis and Results}

%%%
% Figures 1
%%%

\begin{figure*}
  \begin{center}
    \FigureFile(70mm,80mm){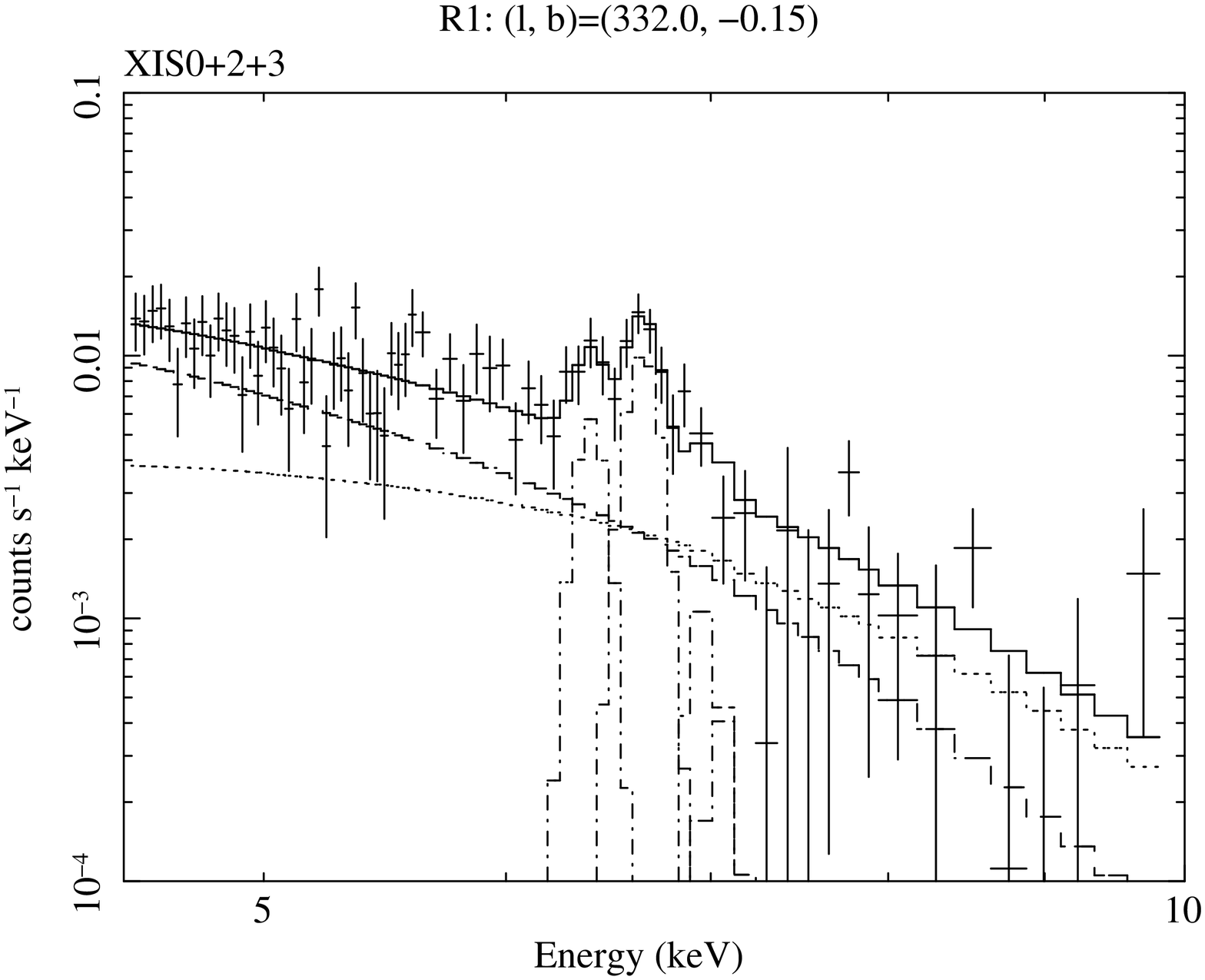}
    \FigureFile(70mm,80mm){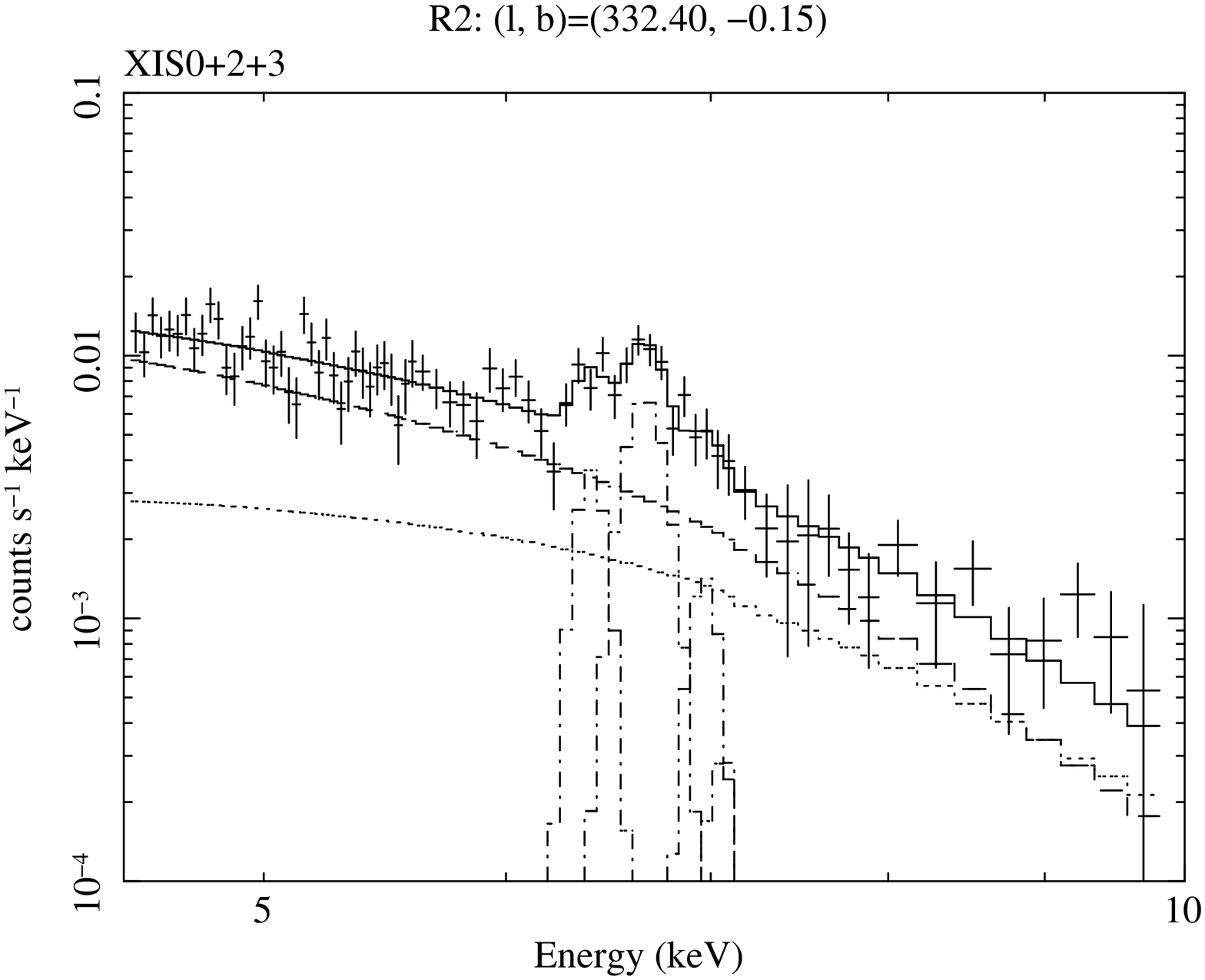}
    \FigureFile(70mm,80mm){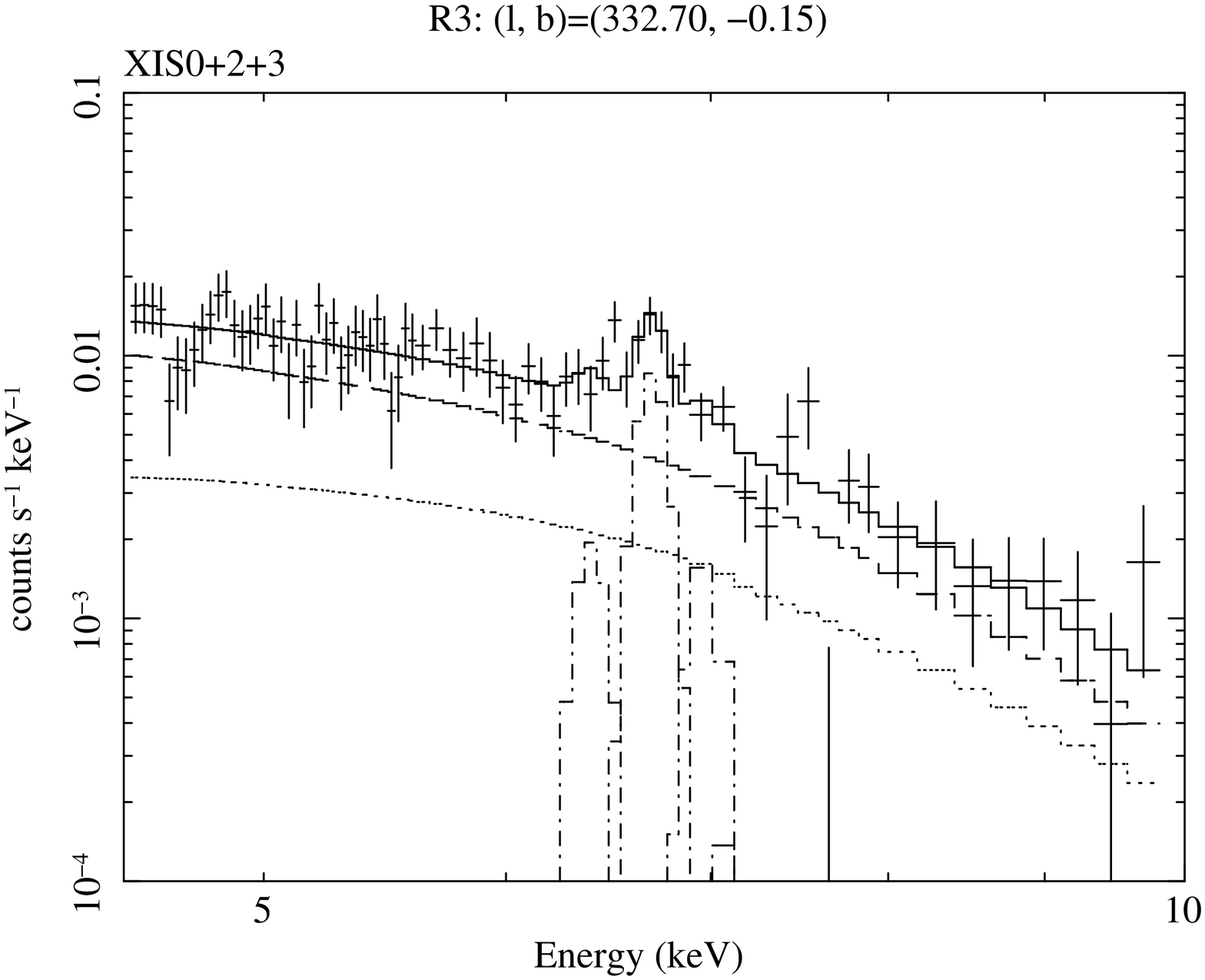}
    \FigureFile(70mm,80mm){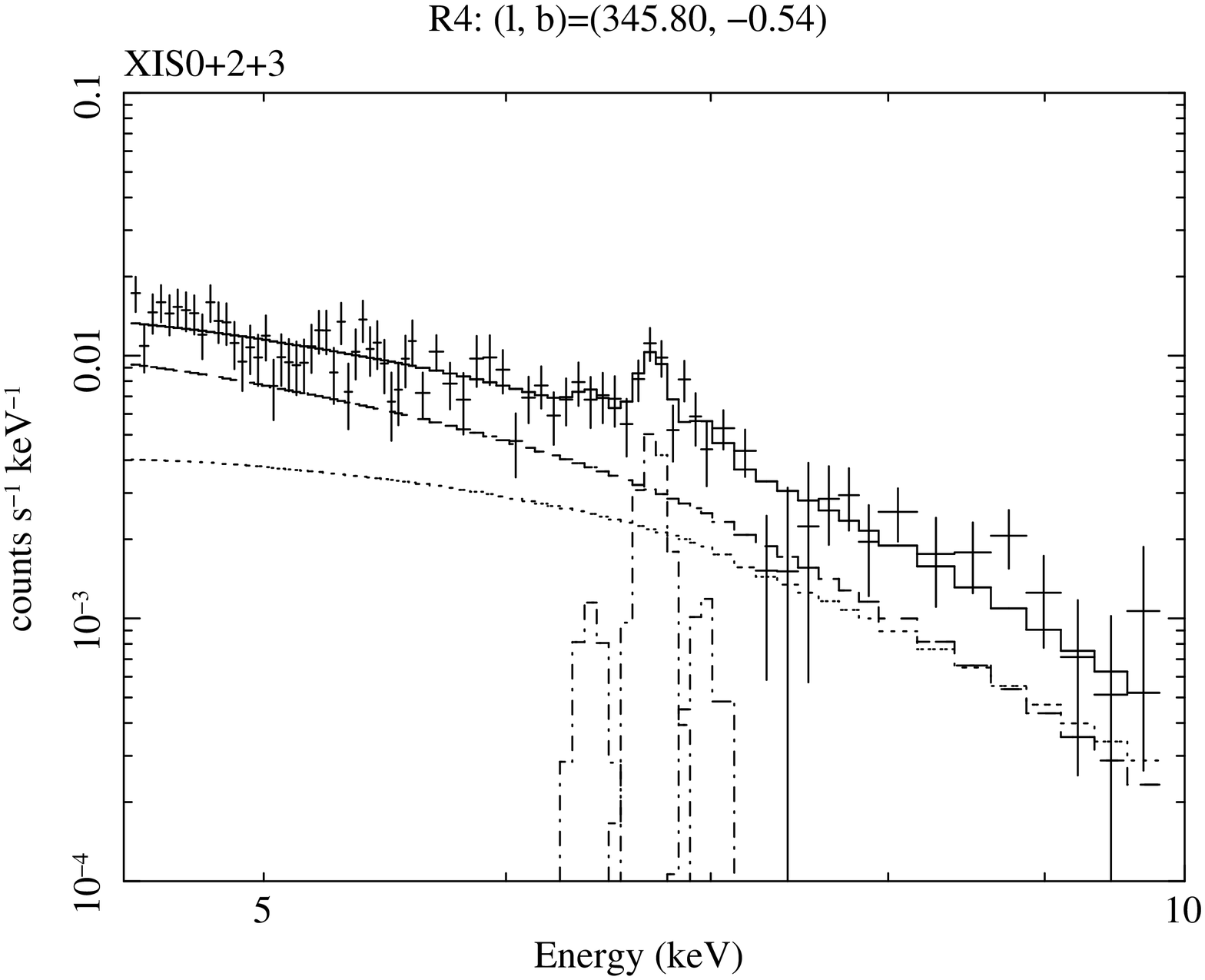}
    \FigureFile(70mm,80mm){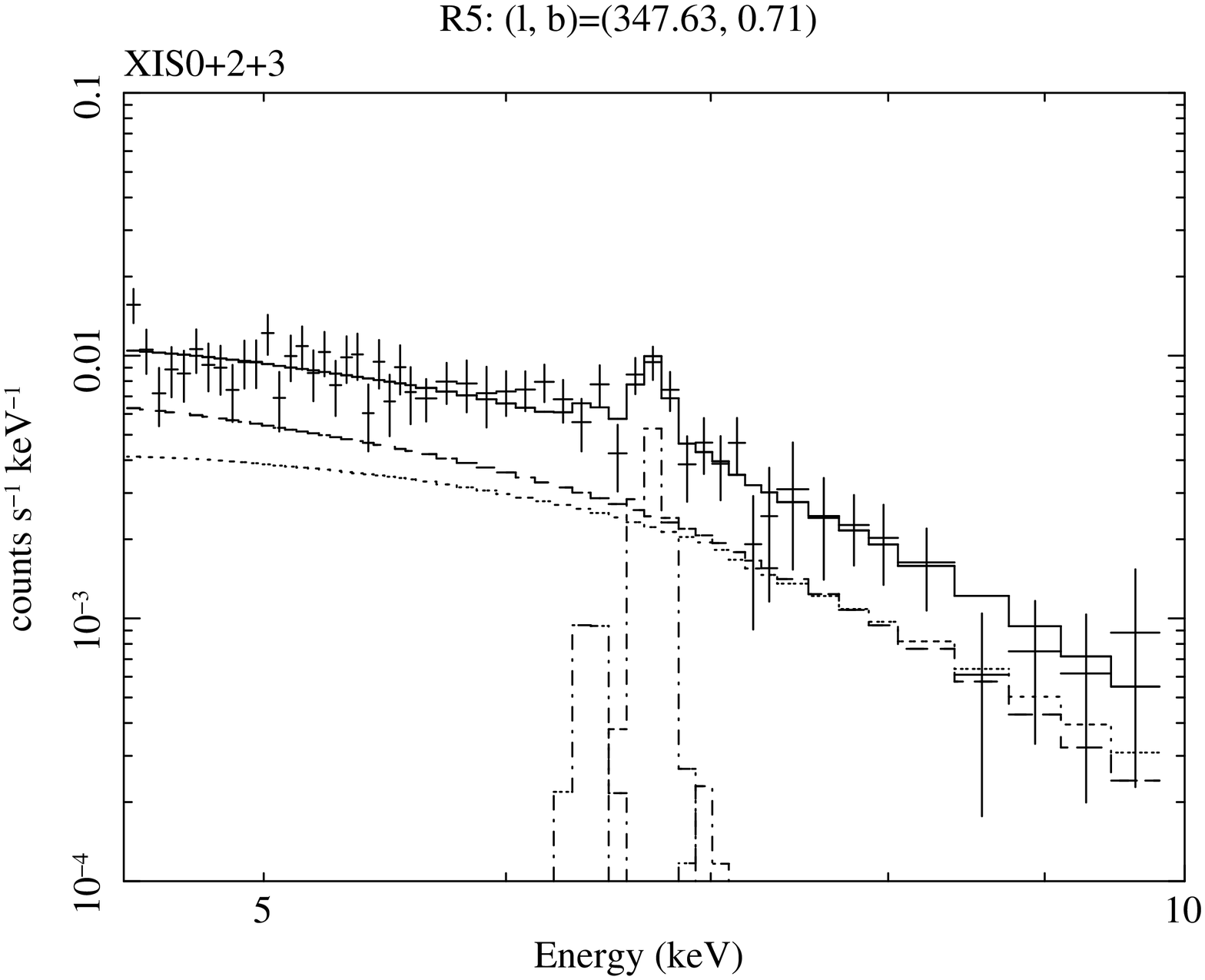}
    \FigureFile(70mm,80mm){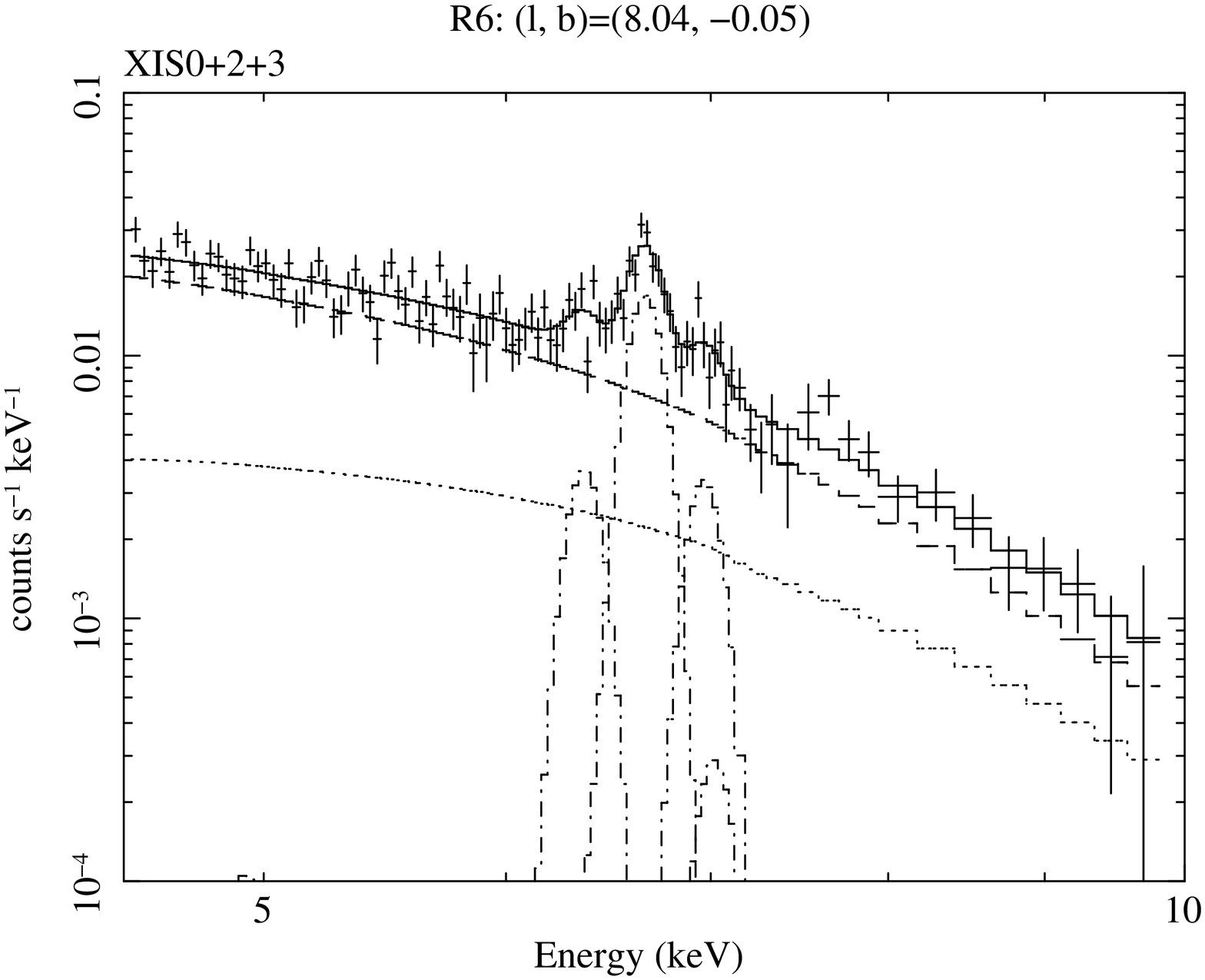}
    \FigureFile(70mm,80mm){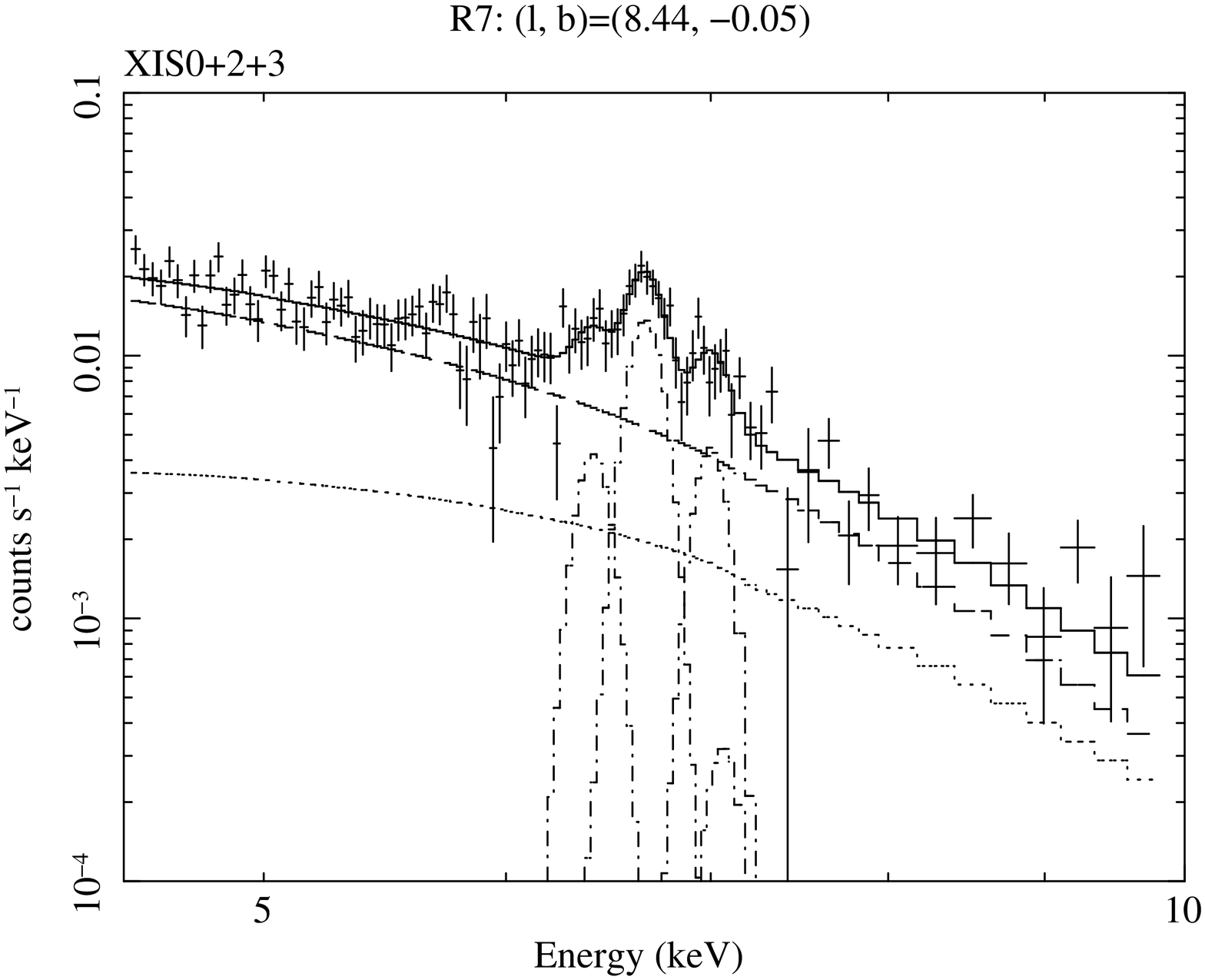}
    \FigureFile(70mm,80mm){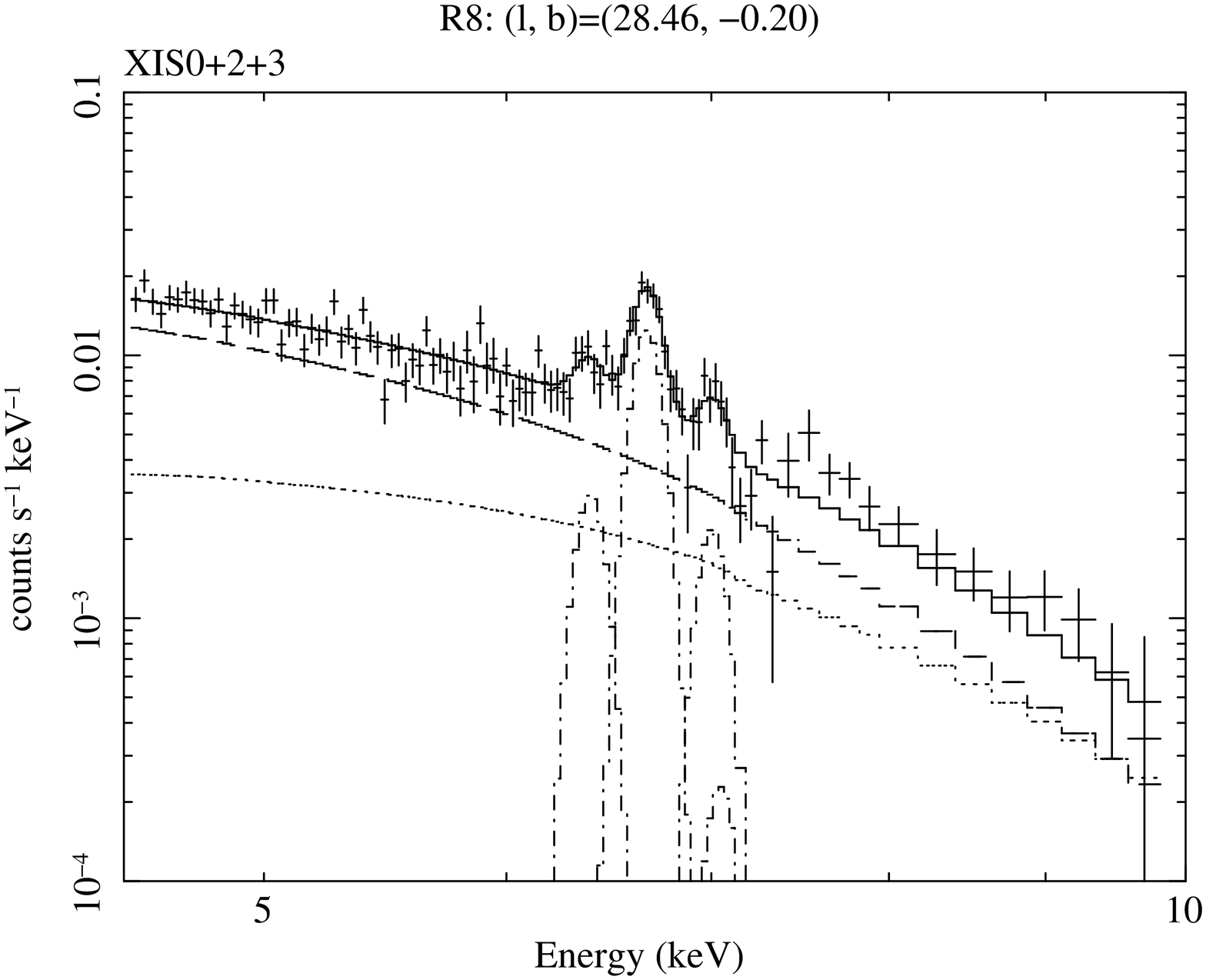}
  \end{center}
  \caption{
The background-subtracted the 
GRXE spectra observed in the various regions (see table 1).
The data obtained with XIS0, 2, and 3 were combined.
The dashed, dotted, and dash-dotted lines show contributions 
of the bremsstrahlung, power-law (CXB), and emission line models, respectively.
}
\label{fig:sample}
\end{figure*}

%%%
% Figures 2
%%%

\begin{figure*}
  \begin{center}
    \FigureFile(70mm,80mm){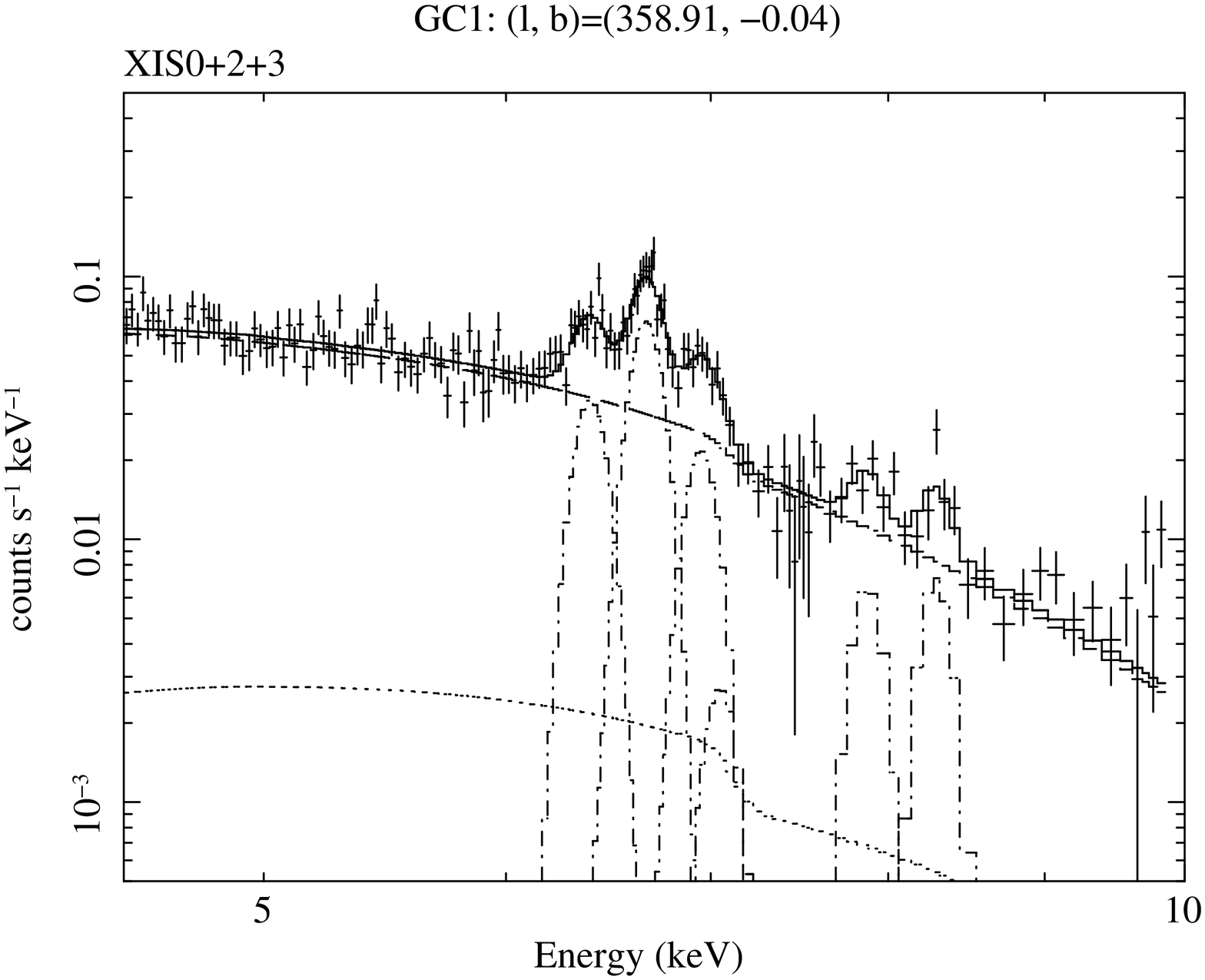}
    \FigureFile(70mm,80mm){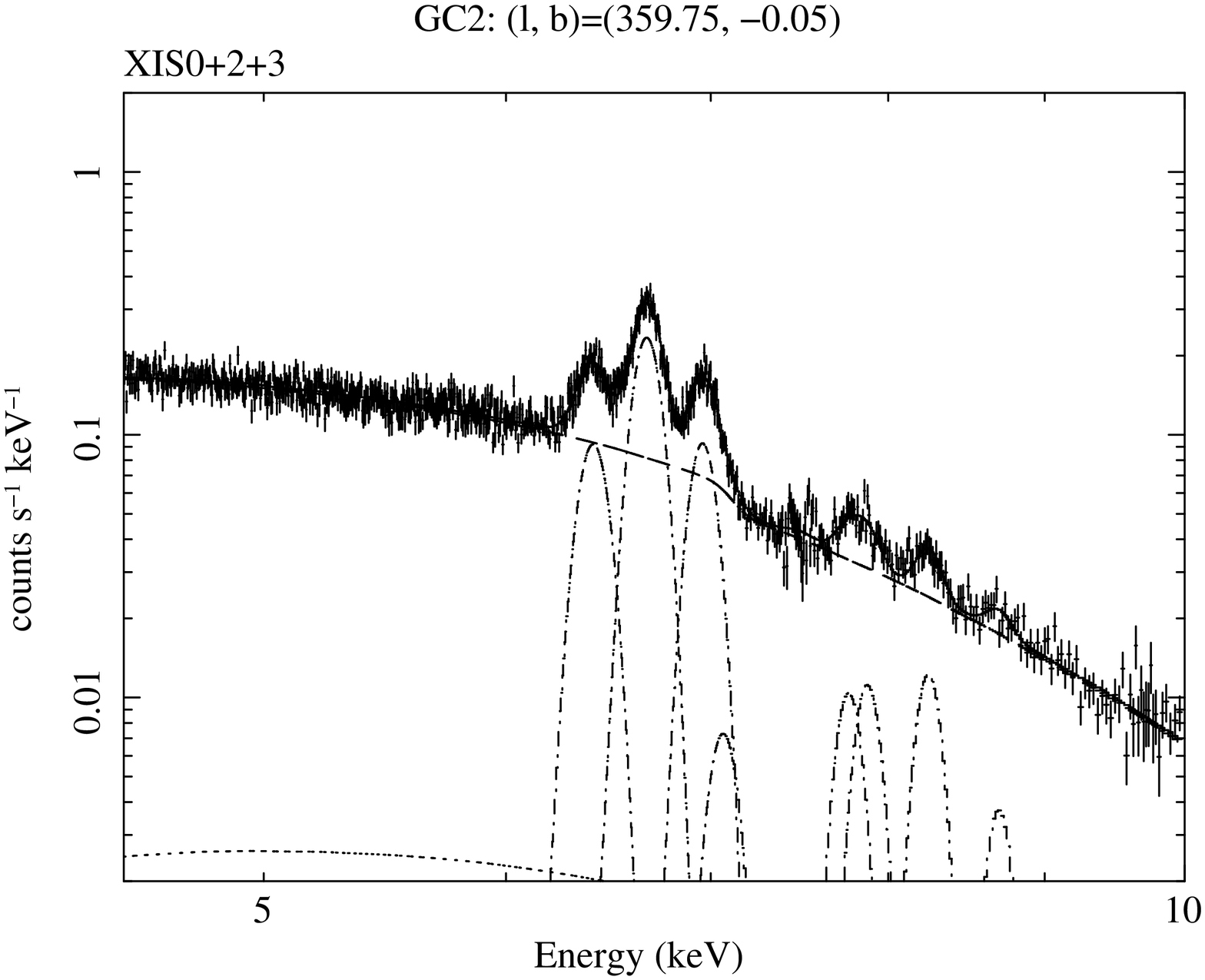}
    \FigureFile(70mm,80mm){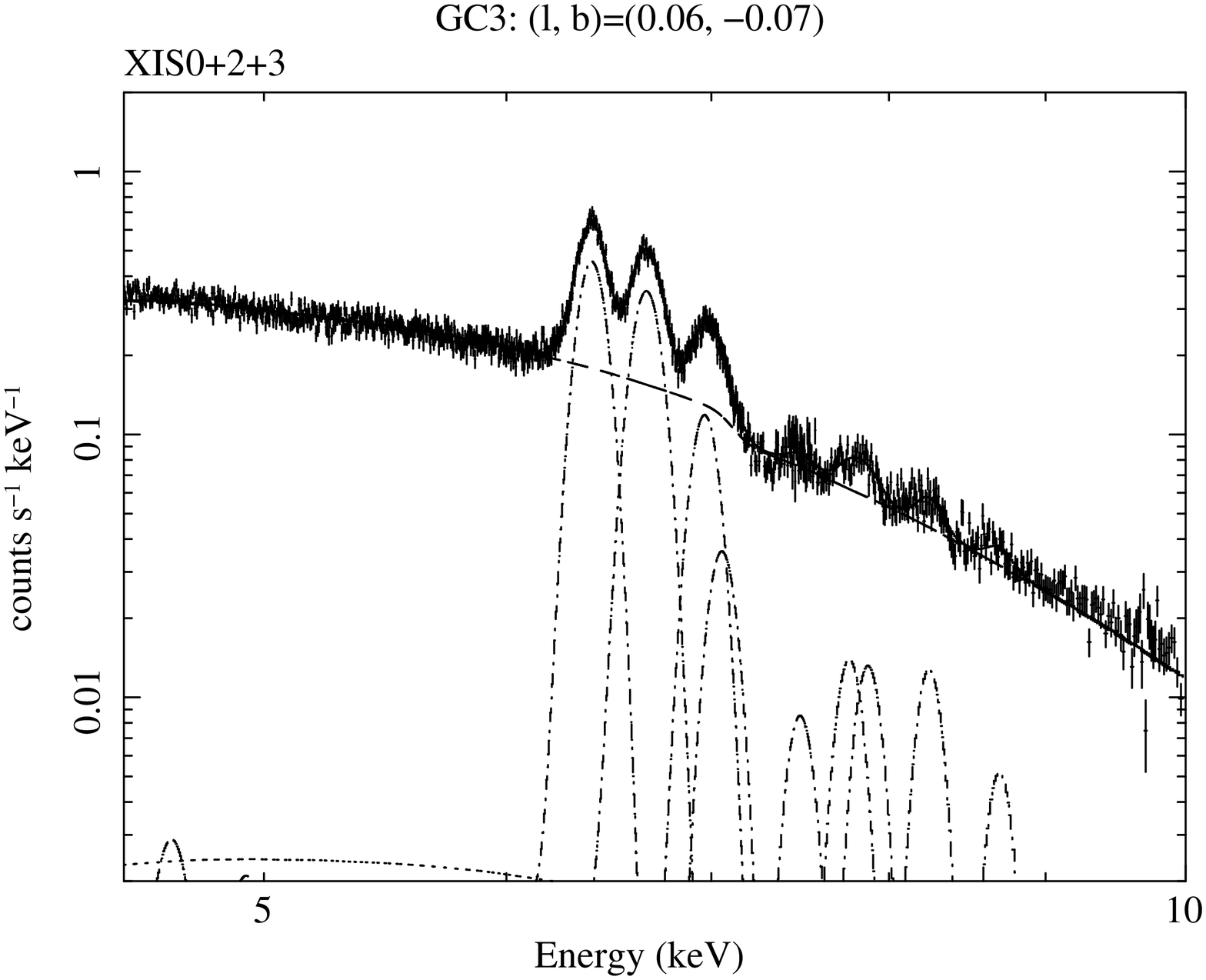}
    \FigureFile(70mm,80mm){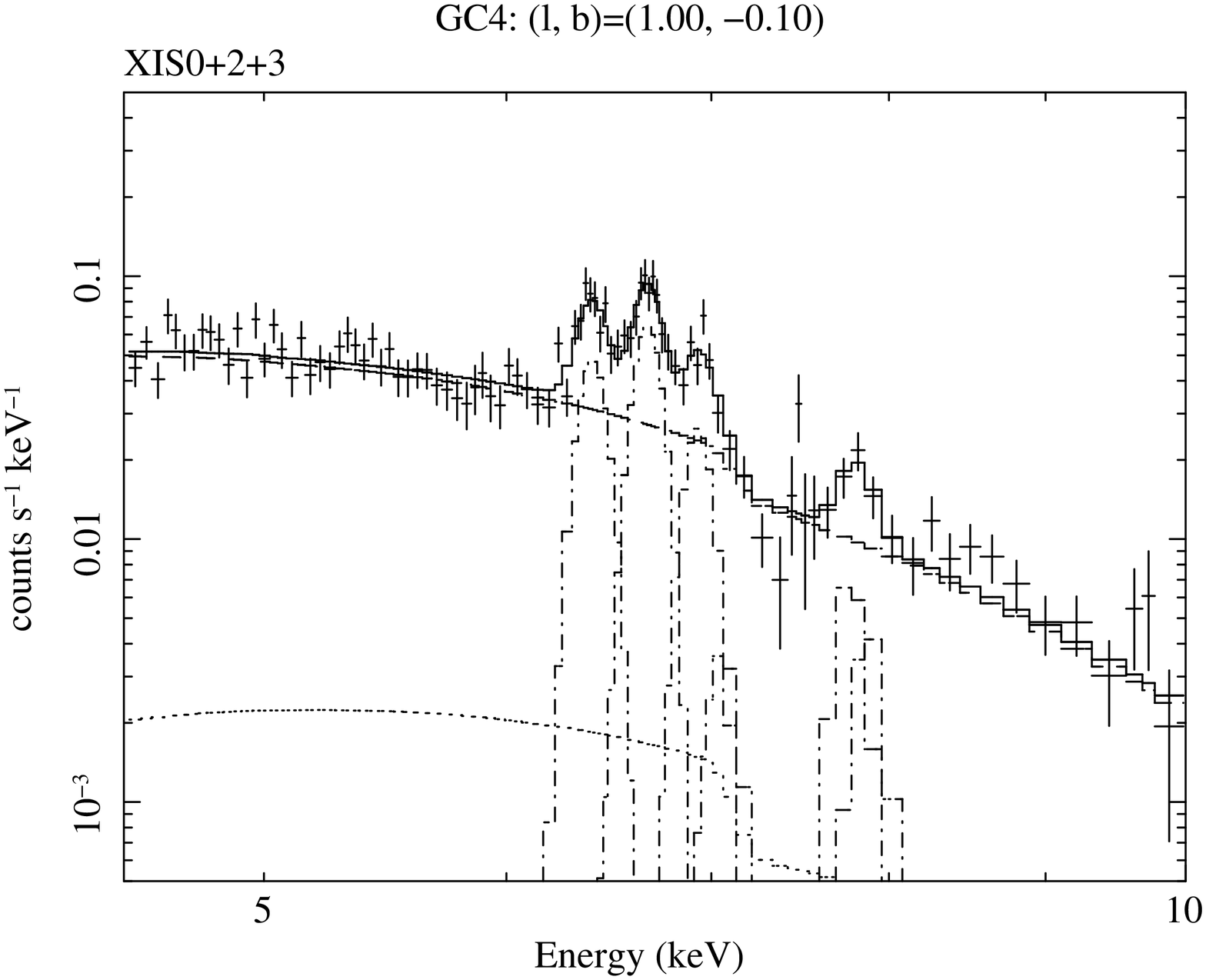}
  \end{center}
  \caption{
The same as figure 1, but the
GCDX spectra (see table 1).
}
\label{fig:sample}
\end{figure*}

In order to construct the GRXE spectra, 
we extracted X-ray photons from 
the entire region of the XIS FOV,
excluding the point-like sources in the FOV and the calibration sources 
located at  corners of the XIS sensors.
We also excluded the Sgr A complex 
(Sgr A$^{\ast}$ and Sgr A East) 
from the GC2 and GC3 data.
To maximize the photon statistics, 
data obtained with XIS0, 2, and 3 
were combined.
Furthermore,  data obtained at the same pointing positions 
(GC2a\&GC2b and GC3a\&GC3b) were also 
combined.
The response files, Redistribution Matrix Files (RMFs)
and Ancillary Response Files (ARFs), were made for each pointing
using {\tt xisrmfgen} and {\tt xissimarfgen}, respectively,  
in HEADAS software version 6.2.
In making ARF for the diffuse GRXE, 
flat photon distribution with a radius of 20$'$ was assumed.
Non-X-ray background was calculated using the background database
provided by the XIS team \citep{Tawa2008}.

In the observed Suzaku spectra, we found many emission lines from 
highly ionized ions, such as Mg, Si, S, and Fe, while
in the present paper we focus on the Fe line feature. 
Figure 1 shows the observed FI spectra in the 4.5--10.0 keV energy band 
after subtraction of the non-X-ray background. 
We can clearly see 
three emission lines in the 6--7 keV range, 
which are   K-lines from neutral (or low ionized), 
He-like, and H-like iron. 
In order to measure the  emission line parameters, 
we fitted the spectra with the following model; 

\medskip 

\noindent For the GRXE

Absorption$\times$(thermal bremsstrahlung$+$Gaussian lines) 

\smallskip

\noindent For the Cosmic X-ray Background (CXB)

Absorption$\times$power-law ($\Gamma$ and $N_{\rm PL}$; fixed) 

\medskip

\noindent 
The cross sections of photoelectric absorption 
are taken from Balucinska-Church and McCammon (1992).
The normalization ($N_{\rm PL}$) and photon index ($\Gamma$) 
of the power-law model was fixed to be 
10 photons s$^{-1}$ cm$^{-2}$ sr$^{-1}$ keV$^{-1}$ at 1 keV 
(e.g., \cite{Gendreau1995,Revnivtsev2005})
and 1.4 \citep{Marshall1980}, 
respectively 
Absorption column densities for the GRXE and the CXB are fixed at
$N_{\rm H}$=3$\times$10$^{22}$ cm$^{-2}$ 
and 6$\times$10$^{22}$ cm$^{-2}$, respectively \citep{Ebisawa2001}.
Three emission lines
of  Fe I K${\alpha}$ (6.4 keV),
Fe XXV K${\alpha}$ (6.7 keV), and Fe XXVI K${\alpha}$ (6.97 keV) 
are considered in the model fitting.
In addition, K${\beta}$ line from Fe I is considered,
whose center energy and intensity are assumed to be
 1.103 $\times$ that of Fe I K${\alpha}$ and 
 0.1 $\times$ that of Fe I K${\alpha}$, 
since they are not constrained from the data.
The best-fit model is also plotted in figure 1 
and the best-fit parameters of the Fe-lines are listed in table 2.
The center energies of three  emission lines have been constrained as
6.36--6.41, 6.65--6.71, and 6.97--7.01 keV.
Thus, Suzaku clearly resolved the Fe line complex into K-lines from Fe I, 
Fe XXV, and Fe XXVI, for all the Galactic plane regions.
No significant broadening in the Fe line width 
was found.

\begin{table*}[tb]
\begin{center}
\caption{Parameters obtained from a spectral analysis for the spectra in the 4.5--10.0 keV.$^{\ast}$}
\small
\begin{tabular}{lcccccccc}
\hline \\  [-6pt]
Parameter & \multicolumn{8}{c}{Value} \\
\hline  \\ [-6pt]
           & R1 & R2 & R3 & R4 & R5 & R6 & R7 & R8 \\
\hline  \\ [-6pt]
$kT^{\dag}$  & 2.6$^{+1.8}_{-0.9}$ & 3.9$^{+1.8}_{-1.1}$ & 10.8$^{+36.0}_{-5.3}$ & 5.4$^{+5.2}_{-2.1}$ & 8.6$^{+35.6}_{-4.4}$ & 6.0$^{+2.0}_{-1.4}$ & 5.1$^{+2.1}_{-1.3}$ & 4.1$^{+1.2}_{-0.8}$\\ 
%%\\ [-4pt]
$E_{\rm 6.4 keV}^{\ddag}$ & 
6.40 (fixed) & 6.40 (fixed) & 6.40 (fixed) & 6.40 (fixed) & 6.40 (fixed) & 6.37$\pm$0.06 & 6.41$^{+0.07}_{-0.06}$ & 6.39$^{+0.05}_{-0.03}$\\
$\sigma_{\rm 6.4 keV}^{\S}$ &
0 (fixed) & 0 (fixed) & 0 (fixed) & 0 (fixed) & 0 (fixed) & $<$133 & 0 (fixed) & $<$84\\
$I_{\rm 6.4 keV}^{\Vert}$ &
2.5$\pm$1.3 & 2.2$^{+1.1}_{-1.2}$ & $<$2.2 & $<$1.3 & $<$1.4 & 1.8$\pm$1.0 & 2.3$\pm$1.2 & 1.4$^{+0.7}_{-0.6}$\\
$EW_{\rm 6.4 keV}^{\sharp}$ &
390$\pm$200 & 190$^{+90}_{-100}$ & $<$160 & $<$150 & $<$190 & 90$\pm$50 & 140$^{+70}_{-80}$ & 110$^{+60}_{-50}$\\ 
%%\\ [-4pt]
$E_{\rm 6.7 keV}^{\ddag}$ & 
6.65$^{+0.04}_{-0.03}$ & 6.67$^{+0.03}_{-0.04}$ & 6.70$\pm$0.04 & 6.71$^{+0.03}_{-0.04}$ & 6.70$\pm$0.04 & 6.666$^{+0.014}_{-0.013}$ & 6.668$\pm$0.017 & 6.674$^{+0.009}_{-0.012}$\\
$\sigma_{\rm 6.7 keV}^{\S}$ &
$<$103 & 65$^{+54}_{-57}$ & 0 (fixed) & $<$77 & $<$78 & $<$38 & $<$65 & $<$36\\
$I_{\rm 6.7 keV}^{\Vert}$ &
5.5$^{+1.9}_{-1.6}$ & 6.2$^{+1.7}_{-1.6}$ & 4.8$^{+1.4}_{-1.5}$ & 2.4$\pm$0.9 & 2.5$^{+1.0}_{-0.9}$ & 9.1$\pm$1.2 & 8.7$^{+1.7}_{-1.5}$ & 6.6$\pm$0.8\\
$EW_{\rm 6.7 keV}^{\sharp}$ &
980$^{+330}_{-280}$ & 600$^{+170}_{-150}$ & 380$\pm$120 & 300$\pm$110 & 400$^{+160}_{-150}$ & 510$\pm$70 & 570$\pm$100 & 590$\pm$70\\ 
%%\\ [-4pt]
$E_{\rm 6.97 keV}^{\ddag}$ &  
6.97 (fixed) & 6.97 (fixed) & 6.97 (fixed) & 6.97 (fixed) & 6.97 (fixed) & 6.97$\pm$0.06 & 7.00$^{+0.07}_{-0.05}$ & 7.01$^{+0.03}_{-0.04}$\\
$\sigma_{\rm 6.97 keV}^{\S}$ & 
0 (fixed) & 0 (fixed) & 0 (fixed) & 0 (fixed) & 0 (fixed) & $<$109 & 0 (fixed) & $<$45\\
$I_{\rm 6.97 keV}^{\Vert}$ & 
$<$1.8 & 1.1$\pm$1.0 & $<$2.4 & $<$1.5 & $<$1.0 & 2.1$^{+1.0}_{-1.1}$ & 3.1$^{+1.3}_{-1.2}$ & 1.3$\pm$0.7\\
$EW_{\rm 6.97 keV}^{\sharp}$ &
$<$390 & 120$\pm$110 & $<$210 & $<$210 & $<$170 & 130$\pm$70 & 230$\pm$90 & 140$\pm$70\\
%%\\ [-4pt]
$\chi^2_{\nu}$ (d.o.f.) & 0.95 (69) & 0.97 (71) & 0.96 (70) & 0.90 (70) & 0.74 (47) & 1.20 (95) & 1.14 (97) & 1.19 (95) \\
\hline\\ [-6pt]
           & \multicolumn{2}{c}{GC1} & \multicolumn{2}{c}{GC2} & \multicolumn{2}{c}{GC3} & \multicolumn{2}{c}{GC4}  \\
\hline  \\ [-6pt]
$kT^{\dag}$  &  \multicolumn{2}{c}{11.7$^{+7.2}_{-3.3}$} &  \multicolumn{2}{c}{13.0$^{+1.5}_{-1.0}$} & \multicolumn{2}{c}{10.8$\pm$0.6} & \multicolumn{2}{c}{18.0$^{+24.2}_{-7.8}$}\\ 
%%\\ [-4pt]
$E_{\rm 6.4 keV}^{\ddag}$ & 
\multicolumn{2}{c}{6.397$^{+0.019}_{-0.020}$} & \multicolumn{2}{c}{6.411$^{+0.006}_{-0.005}$} & \multicolumn{2}{c}{6.402$\pm$0.002} & \multicolumn{2}{c}{6.404$^{+0.017}_{-0.013}$} \\
$\sigma_{\rm 6.4 keV}^{\S}$ &
\multicolumn{2}{c}{$<$68} &  \multicolumn{2}{c}{$<$38} & \multicolumn{2}{c}{$<$10} & \multicolumn{2}{c}{$<$46} \\
$I_{\rm 6.4 keV}^{\Vert}$ &
\multicolumn{2}{c}{17$\pm$4} & \multicolumn{2}{c}{45$\pm$3} & \multicolumn{2}{c}{229$^{+4}_{-5}$} & \multicolumn{2}{c}{24$\pm$5} \\
$EW_{\rm 6.4 keV}^{\sharp}$ &
\multicolumn{2}{c}{200$\pm$50} & \multicolumn{2}{c}{180$\pm$20} & \multicolumn{2}{c}{440$\pm$10} & \multicolumn{2}{c}{270$\pm$60} \\ 
%%\\ [-4pt]
$E_{\rm 6.7 keV}^{\ddag}$ & 
\multicolumn{2}{c}{6.673$\pm$0.010} & \multicolumn{2}{c}{6.675$^{+0.003}_{-0.002}$} & \multicolumn{2}{c}{6.671$^{+0.003}_{-0.002}$} & \multicolumn{2}{c}{6.676$\pm$0.012} \\
$\sigma_{\rm 6.7 keV}^{\S}$ &
\multicolumn{2}{c}{$<$28} & \multicolumn{2}{c}{$<$22} & \multicolumn{2}{c}{24$^{+5}_{-6}$} & \multicolumn{2}{c}{$<$46}\\
$I_{\rm 6.7 keV}^{\Vert}$ &
\multicolumn{2}{c}{36$\pm$4} & \multicolumn{2}{c}{119$^{+4}_{-3}$} & \multicolumn{2}{c}{202$^{+5}_{-4}$} & \multicolumn{2}{c}{36$\pm$6}\\
$EW_{\rm 6.7 keV}^{\sharp}$ &
\multicolumn{2}{c}{450$^{+60}_{-50}$} & \multicolumn{2}{c}{510$\pm$20} & \multicolumn{2}{c}{420$^{+20}_{-10}$} & \multicolumn{2}{c}{420$^{+80}_{-70}$} \\ 
%%\\ [-4pt]
$E_{\rm 6.97 keV}^{\ddag}$ & 
\multicolumn{2}{c}{6.95$\pm$0.03} & \multicolumn{2}{c}{6.962$\pm$0.004} & \multicolumn{2}{c}{6.970$^{+0.004}_{-0.005}$} & \multicolumn{2}{c}{6.93$^{+0.03}_{-0.02}$} \\
$\sigma_{\rm 6.97 keV}^{\S}$ & 
\multicolumn{2}{c}{$<$53} & \multicolumn{2}{c}{$<$14} & \multicolumn{2}{c}{$<$16} & \multicolumn{2}{c}{$<$49}\\
$I_{\rm 6.97 keV}^{\Vert}$ & 
\multicolumn{2}{c}{13$\pm$4} & \multicolumn{2}{c}{52$\pm$3} & \multicolumn{2}{c}{72$^{+4}_{-3}$} & \multicolumn{2}{c}{16$\pm$5}\\
$EW_{\rm 6.97 keV}^{\sharp}$ &
\multicolumn{2}{c}{180$\pm$60} & \multicolumn{2}{c}{240$^{+20}_{-10}$} & \multicolumn{2}{c}{160$^{+20}_{-10}$} & \multicolumn{2}{c}{200$\pm$60}\\
%%\\ [-4pt]
$\chi^2_{\nu}$ (d.o.f.) & \multicolumn{2}{c}{1.03 (146)} & \multicolumn{2}{c}{1.03 (853)} & \multicolumn{2}{c}{1.07 (1010)} & \multicolumn{2}{c}{1.11 (85)}\\
\hline \\
\end{tabular}
\end{center}
\vspace{-0.5cm}
$^{\ast}$ Errors show single-parameter 90\% confidence levels.\\
$^{\dag}$ Temperature of the thermal bremsstrahlung model in the unit of keV.\\
$^{\ddag}$ Energy of the emission line in the unit of keV.\\
$^{\S}$ Width of the emission line in the unit of eV.\\
$^{\Vert}$ Intensity of the emission line in the 
unit of $\times10^{-8}$ counts s$^{-1}$ cm$^{-2}$ arcmin$^{-2}$.\\
$^{\sharp}$ Equivalent width in the unit of eV.\\
\normalsize
\end{table*}

In order to obtain the iron line fluxes from the GCDX, 
we applied the same process to the GC data.
We constructed the GCDX spectra excluding the Sgr A complex and
point-like sources in the FOV.
The absorption column for the GCDX was fixed to 
$N_{\rm H}$=6$\times$10$^{22}$ cm$^{-2}$ \citep{Sakano2002} and 
that for CXB was set to be twice of the value for the GCDX; 
$N_{\rm H}$=12$\times$10$^{22}$ cm$^{-2}$.
As noted in \citet{Koyama2007c}, 
a clear absorption edge of neutral or lower ionized iron 
was found at 7.1 keV in the spectra, so
we set the Fe abundance in the absorption column free. In addition,
emission lines of Ni I K$\alpha$ (7.49 keV), Ni XXVII K$\alpha$ (7.77 keV), 
Fe XXV K$\beta$ (7.88 keV), Fe XXVI K$\beta$ (8.25 keV), 
Fe XXV K$\gamma$ (8.29 keV), and 
Fe XXVI K$\gamma$ (8.70 keV) have been found in the GCDX spectra 
\citep{Koyama2007c}.
Therefore, 
we added these emission lines to the model, if they are found in the spectra. 
In this analysis, the line center energy 
and the intrinsic line width were fixed to 
the predicted values and null, respectively.
The best-fit model is plotted in figure 2 and the best-fit parameters 
of Fe-lines are listed in table 2.

Figure 3 shows the 6.4 keV, 6.7 keV, and 6.97 keV emission line fluxes 
as a function of the Galactic longitude, which 
demonstrates a
 significant enhancement of the emission line fluxes at the GC.
The intensities of the Fe-lines in the R4 and R5 regions,
which are $>$\timeform{0.D5} apart from the Galactic plane, are 
systematically lower than those in the other Galactic plane regions 
(R1--3, 6--8), because
the GRXE intensity decreases with the height from the Galactic plane 
\citep{Yamauchi1993, Kaneda1997}.
Taking account of the galactic latitude-dependence of the GRXE intensity,
%%we found that 
the 6.7 keV line distribution is consistent with those 
obtained in previous observations 
\citep{Koyama1989,Yamauchi1993,Sugizaki2001}.
The 6.4 keV line is also found in the various Galactic plane regions 
 ($b\sim0^{\circ}$) and 
particularly conspicuous close to the GC.

Figure 4 shows the flux ratios of the 6.4 keV/6.7 keV and 
6.97 keV/6.7 keV lines.
The 6.4 keV/6.7 keV ratios for the GCDX are significantly 
larger than those for the GRXE and 
the 6.97 keV/6.7 keV ratios also show similar tendency.

%%%
% Figures 3
%%%

\begin{figure}[t]
  \begin{center}
    \FigureFile(80mm,80mm){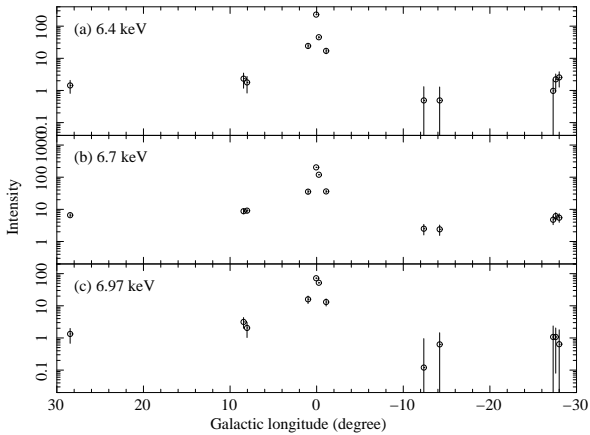}
  \end{center}
  \caption{
(a) 6.4 keV, (b) 6.7 keV, and (c) 6.97 keV line fluxes as a function of 
the Galactic longitude.
The unit of the line flux is $\times10^{-8}$ count s$^{-1}$ cm$^{-2}$ 
arcmin$^{-2}$.
The error shows the 90 \% confidence level.
}
\label{fig:sample}
\end{figure}

%%%
% Figures 4
%%%

\begin{figure}[t]
  \begin{center}
    \FigureFile(80mm,80mm){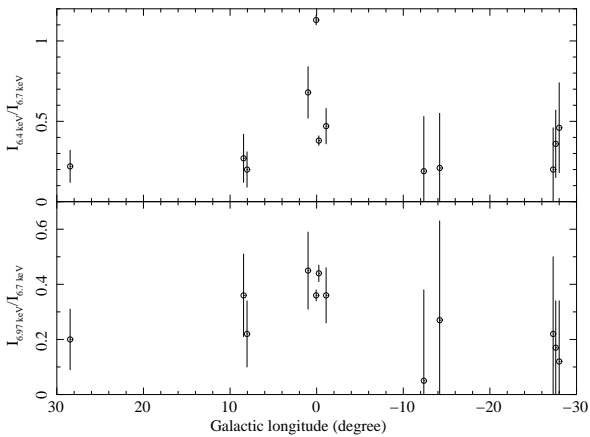}
  \end{center}
  \caption{
The flux ratios of 6.4 keV/6.7 keV (upper) and 
6.97 keV/6.7 keV (lower) lines as a function of the 
Galactic longitude.
The error shows the 90 \% confidence level.
}
\label{fig:sample}
\end{figure}

\section{Discussion}

\subsection{The 6.7 and 6.97 keV Lines}

Thanks to the excellent energy resolution of the Suzaku XIS, 
the Fe line complex was clearly resolved into three narrow emission lines
at $\sim$6.4, $\sim$6.7, and $\sim$6.97 keV, and
their spatial distribution along the Galactic plane was manifested.
The 6.7 keV line is clearly found in all the Galactic Ridge and GC
regions, indicating
that a thin thermal hot plasma %with a temperature of $\sim$5--8 keV 
is located along the Galactic plane.
The equivalent widths (EWs) of the 6.7 keV and 6.97 keV lines 
are $\sim$300--980 eV and $\sim$20--230 eV, respectively.
The observed center energies of the 6.7 keV line are consistent with the 
theoretical value of Fe XXV in a collisional ionization equilibrium 
(CIE) plasma (6.68 keV), taking account of the statistical errors 
and the calibration uncertainty of the energy scale 
(0.2\% at 6 keV, \cite{Koyama2007a}).
Furthermore, the flux ratios of the 6.97 keV/6.7 keV lines 
are in good agreement with those expected 
from a CIE plasma with a temperature 
of 2--8 keV.
Thus, our results support that 
the GRXE is likely to be thermal emission from a CIE plasma, 
as shown for the GCDX by \citet{Koyama2007c} 
and for one of the GRXE fields (R8) by \citet{Ebisawa2008}.

\subsection{The 6.4 keV Line}

In addition to the 6.7 keV and 6.97 keV lines, 
the 6.4 keV emission line was found in various Galactic plane regions,
which suggests omnipresence of the 6.4 keV line in the Galactic plane regions.
The EW of the 6.4 keV line from the GRXE is $<$400 eV.
The GC region has been known to exhibit a strong 6.4 keV line, in particular
strong  enhancement of the 6.4 keV line flux at the Sgr B2 cloud was found
with ASCA \citep{Koyama1996}.
%%Omnipresence of the 6.4 keV line in the Galactic plane regions
%% is revealed for the first time.
The Sgr B2 X-ray spectrum exhibited a strong 6.4 keV emission line 
with the EW
of $>$1 keV and an absorption edge of the neutral iron at 7.1 keV, which
is well explained by reflection 
of X-rays from an external bright X-ray source 
(X-ray Reflection Nebula, XRN), where Sgr A* is presumably
the irradiating source \citep{Sunyaev1993,Koyama1996}.
Recently similar XRN objects have been 
found in the GC region with 
ASCA, Chandra, and Suzaku 
(Murakami et al. 2000, 2001a, 2001b; \cite{Park2004},
Koyama et al. 2007b, 2008; \cite{Nobukawa2008}).
On the other hand,  
the GRXE fields are much far from the GC, 
thus
the same XRN scenario as that in the GC region cannot be expected.
Therefore, another scenario
to produce the 6.4 keV emission line is required.
Interactions between the interstellar medium and
high-energy electrons or X-ray photons in the Galactic plane fields
can produce the 6.4 keV lines.
Assuming the diffuse origin of the GRXE,
\citet{Valinia2000} proposed that interactions between 
the interstellar medium and cosmic-ray electrons 
are responsible for the 6.4 keV line.
In this case,
distribution of the 6.4 keV line intensity is expected to be spatially 
correlated with that of the molecular cloud, which may be observable in future.

If the GRXE is composition of numerous faint X-ray sources, 
the composite spectra
must exhibit a similar 
spectrum to the GRXE, in particular the 
three narrow Fe emission lines at the energies of 6.4, 6.7, and 6.97 keV. 
Cataclysmic variables (CVs) and active binary stars (ABs) 
such as RS CVn type stars are 
proposed to be prime candidate populations of the GRXE \citep{Revnivtsev2006}.
The X-ray spectrum of CVs
exhibits a thermal emission with the 6.4, 6.7, and 6.97 keV 
lines (e.g., \cite{Hellier1998,Ezuka1999,Ishida2007,Mukai2007}), 
while that of ABs exhibits a thermal emission with the 6.7 keV line 
but without the strong 6.4 keV line
(e.g., \cite{Gudel1999}).
Consequently, the Fe line features of the GRXE,
omnipresence of the 6.4 keV line in addition to the 6.7 keV and 6.97 keV 
lines, implies that 
CVs are major contributor to the flux of the GRXE, 
if the GRXE has the point source origin.

\subsection{Fe Lines from CVs }

Magnetic CVs (polars and intermediate polars), a subclass of CVs,
are efficient X-ray emitters with luminosities of 
$\sim$10$^{32}$--10$^{33}$ erg s$^{-1}$, while 
the majority of CVs are non magnetic CVs, faint X-ray sources 
with luminosities of $\sim$10$^{31}$ erg s$^{-1}$. 
At least 80\% of non magnetic CVs are thought to be 
dwarf novae (e.g., \cite{Patterson1984}).
According to Fe line properties of magnetic CVs 
summarized in \citet{Hellier1998} and Ezuka and Ishida (1999),
the mean EWs of 
the 6.4 keV, 6.7 keV, and 6.97 keV lines are $\sim$100--150 eV, $\sim$200 eV, 
and $\sim$100 eV, respectively.
Suzaku results of dwarf novae, SS Cyg and V893 Sco \citep{Ishida2007,Mukai2007}
suggest that their Fe line features are similar to those of magnetic CVs.
Comparing the Fe line features of the GRXE with those of CVs,
we found that 
the 6.7 keV line EW of the GRXE ($\sim$300--980 eV) is larger than that 
of CVs, 
but the Fe line properties 
are similar to each other.

We estimate the space density of CVs which are required 
to explain all the observed Fe line flux of the GRXE.
If the continuum emission of CVs is thin thermal emission from a hot plasma 
with a temperature of $\sim$10 keV and the EW of the 6.4 keV line is 150 eV, 
the luminosity of the 6.4 keV line emission is estimated to be 
$\sim$1.5\% of the 2--10 keV luminosity.
The longitudinal distribution of the GRXE shows 
a strong enhancement within the area of $|l|\le$30$^{\circ}$
\citep{Yamauchi1993,Sugizaki2001,Revnivtsev2006},
which corresponds to the Galactic inner disk with a radius of 4 kpc.
For simplicity, we assumed 
that the CVs are uniformly 
distributed in the Galactic inner disk region 
with a radius ($R_{\rm d}$) of 4 kpc and a height of 300 pc and 
the distance to the Galactic Center ($R_0$) is 8 kpc. 
Based on the results of X-ray observations of CVs 
(e.g., \cite{Cordova1984,Mukai1993,Ezuka1999,Baskill2005,Sazonov2006}),
we also assumed the mean luminosity to be 10$^{32}$ erg s$^{-1}$. 
Thus, the observed 6.4 keV line flux from the GRXE, 
$F_{\rm 6.4 keV}$ (erg s$^{-1}$ cm$^{-2}$ str$^{-1}$), is expressed as follows.
\begin{equation}
 F_{\rm 6.4 keV} = \int \frac{n\ L_{\rm 6.4 keV}}{4\pi} dx = \frac{n\ L_{\rm 6.4 keV}}{4\pi} X,
\end{equation}
where $L_{\rm 6.4 keV}$, $n$, and $X$ are a mean 6.4 keV line luminosity 
of CV (= 1.5\% of 10$^{32}$ erg s$^{-1}$), 
a space density of CV, and a length across the disk, respectively.
In the case of the uniform disk model, $X$ is 
2$\sqrt{R_{\rm d}^2 - (R_0 {\rm sin}\ l)^2}$. 
Using the best-fit 6.4 keV line surface brightness, 
the required CV space density is calculated to be 
$\sim$(0.6--8.6)$\times$10$^{-5}$ pc$^{-3}$.

We also carried out the same estimation using the 6.7 keV line.
If we assume the mean EW of the 6.7 keV line from CVs to be 200 eV,
the 6.7 keV line luminosity is estimated to be $\sim$2 \%
of the 2--10 keV luminosity.
Then, using the observed Fe line surface brightness from the GRXE, 
the required CV space density is calculated to be 
(2.6--21)$\times$10$^{-5}$ pc$^{-3}$.

The estimated CV space densities are %%significantly 
larger than 
that of CVs observed in the Solar neighborhood so far 
($\sim6\times$10$^{-6}$ pc$^{-3}$, \cite{Patterson1984}).
However, taking account of uncertainties of the current estimation of the 
CV space density in the Galaxy,
we cannot completely rule out the possibility 
that a large number of CVs are hiding inside the Galaxy.
Or, if the mean luminosity of CVs turns out to be significantly larger than 
10$^{32}$ erg s$^{-1}$, the required CV space density can be much lower.
For more precise estimation, 
construction of the luminosity function 
and a realistic model of the source distribution in the Galaxy 
are needed.
Detailed measurements of the Fe line features of CVs are also useful.
In any case, if the CV origin is correct,
the reflection profile by the Compton scattering and 
the line broadening are expected,
because the 6.4 keV emission line of CVs is emitted from 
the white dwarf surface and/or the accreting matter.
A systematic energy shift of $\sim$2 eV by the gravitational redshift 
is also expected.
These features would be revealed by future X-ray micro-calorimeter missions.

\subsection{The Flux Ratios of the Fe Lines}

The 6.7 keV line is clearly found in all the GRXE and GCDX spectra, while
the intensities of the $\sim$6.4 keV and 
$\sim$7 keV lines relative to 
that of the $\sim$6.7 keV line seems to vary from field to field 
(figures 1 and 2).
Figure 4 shows that there are some differences in the line flux ratios,
although the errors are large.
In particular, the line flux ratios of the GCDX are systematically larger than 
those of the GRXE.
The 6.4 keV line would originate from a non thermal process.
Here, we focus on the flux ratios of the other two lines in the thermal 
origin, the 6.97 keV/6.7 keV lines.

In order to examine the hypothesis that the line flux ratios 
observed at all the positions are the same,
we fitted all the 6.97 keV/6.7 keV line flux ratios (GC1--4 and R1--8) 
with a constant model 
and obtained the reduced $\chi^2$ value of 2.90 (the degree of freedom of 11).
This means that the hypothesis is statistically rejected.
Furthermore, 
the weighted mean values of the 6.97 keV/6.7 keV line flux ratio of the GCDX
and the GRXE are 0.38$\pm$0.02 and 0.22$\pm$0.06
(90\% confidence level), respectively.
Thus, %%we conclude that at least 
the line flux ratio of the GRXE is 
not the same as that of the GCDX.
Based on the APEC model in XSPEC, 
the mean ratio for the GCDX (0.38) is corresponding to 
$\sim$7 keV, while 
that for the GRXE (0.22) is $\sim$5.5 keV.
We also applied the same analysis to only the GRXE data (R1--8) and obtained 
the reduced $\chi^2$ value of 0.49 (the degree of freedom of 7).
From the statistical point of view,
all the line flux ratio of the GRXE is considered to be
the same.
To examine the difference in the line ratios for the GRXE,
we need good statistics data and more samples of the GRXE.

If the GRXE and the GCDX are composition of numerous faint X-ray sources, 
the spectra should be an averaged spectrum of the
X-ray sources contributing to the GRXE and the GCDX.
According to the results of the fluctuation analysis of the 
ASCA data \citep{Sugizaki1999},
more than 10$^{3}$ faint sources responsible for the GRXE should be included 
in the XIS FOV.
In the case of the GCDX, the number of contributing sources should be 
larger because the 6.7 keV and 6.97 keV line fluxes 
of the GCDX are more than 10 times larger than those of the GRXE.
Hence, the GRXE and the GCDX spectra observed in the various regions 
are expected to be quite similar to each other. 
The difference in the 6.97 keV/6.7 keV line flux ratio suggests that 
at least the candidate sources in the GC region have a 
higher temperature plasma emission than those on the Galactic plane 
systematically.

On the other hand, if the GRXE has the 
diffuse origin, the difference in the line flux ratio of 
6.97 keV/6.7 keV can be explained by spatial difference of the plasma 
temperature.
If the hot plasma is produced by a process concerning the activity 
in the Galactic scale, 
the temperature distribution may be correlated with the Galactic structure. 
For example, the temperatures in the central parts are expected to be higher, 
while those in the outer regions and the off-plane regions 
are relatively low, 
which qualitatively agrees with the observation.
More precise systematic study of the spatial distribution 
of the Fe lines is expected to give the answer in future.

\section{Summary}

We analyzed Suzaku data taken at various regions along the 
Galactic plane and studied their Fe-K emission line features. 
Suzaku resolved the Fe line complex into three narrow lines 
at $\sim$6.4 keV, $\sim$6.7 keV and $\sim$6.97 keV, 
which are K-lines from neutral (or low-ionized), 
He-like, and H-like iron ions, respectively. 
The 6.7 keV line is clearly seen in all the observed regions
and the 6.4 keV emission line was also found in various 
Galactic plane regions.
The 6.4 keV/6.7 keV and the 6.97 keV/6.7 keV line flux ratios of 
the GCDX are found to be systematically larger than those of the GRXE.

\vspace{1pc}

Authors are grateful to all the members of the Suzaku team.
This work was supported in part by the Grant-in-Aid for Scientific Research
of the Japan Society for the Promotion of Science (JSPS)
(No. 18540228, S.Y.).

%%%
% See the manual for the detail.
%%%

\end{document}